\documentclass[12pt]{iopart}

\usepackage{graphicx}
\usepackage{bm}
\usepackage{iopams}

\newcommand{\myscale}{0.9}

\newcommand{\myreffig}[1]{figure \ref{#1}}
\newcommand{\myrefeq}[1]{(\ref{#1})}
\newcommand{\Hfx}[1]{Hf$_{100-x}$#1$_x$}
\newcommand{\HfCux}{\Hfx{Cu}}
\newcommand{\HfFex}{\Hfx{Fe}}
\newcommand{\TiCu}{Ti$_{65}$Cu$_{35}$}

\newcommand{\Ng}{$\mNg$}
\newcommand{\NE}{$\mNE$}
\newcommand{\Tc}{$\mTc$}
\newcommand{\Hc}{$\mHc$}
\newcommand{\HcT}{$\mHcT$}

\newcommand{\mNg}{N_{\gamma}(E_F)}
\newcommand{\mNE}{N(E_F)}
\newcommand{\dHdT}{(dH_{c2}/dT)_{\mTc}}
\newcommand{\lso}{\lambda_{\mathrm{so}}}
\newcommand{\lep}{\lambda_{\mathrm{ep}}}
\newcommand{\lesf}{\lambda_{\mathrm{esf}}}
\newcommand{\mTc}{T_c}
\newcommand{\mHc}{H_{c2}}
\newcommand{\mHcT}{\mHc(T)}

\begin{document}

\title{Enhanced superconductivity in Hf--base metallic glasses}

\author{Emil Tafra$^1$, Mario Basleti\'{c}$^1$, Ramir Risti\'{c}$^2$, Emil Babi\'{c}$^1$ and Amir Hamzi\'{c}$^1$}
\address{$^1$Department of Physics, Faculty of Science, P.O.Box 331, HR-10002 Zagreb, Croatia}
\address{$^2$Department of Physics, University of Osijek, P.O.Box 125, HR-31000 Osijek, Croatia}

\date{\today}

\begin{abstract}
Systematic study of electrical resistivity of Hf$_{100-x}$Fe$_x$
$(x=20, 25)$, Hf$_{100-x}$Cu$_x$ $(x=30,40,50)$, and
Ti$_{65}$Cu$_{35}$ metallic glasses has been done in the
temperature range $0.3\,\mathrm{K}-290\,$K, and in magnetic fields
$B\le 5\,$T. All Hf--base alloys are superconducting with $T_c \ge
0.44\,$K, which is well above $T_c$ of pure crystalline Hf
($0.13\,$K). From the initial slopes of the upper critical fields,
$(dH_{c2}/dT)_{T_c}$, and resistivities we determined the dressed
electronic densities of states, $N_{\gamma}(E_F)$, for all alloys.
Both $T_c$ and $N_{\gamma}(E_F)$ decrease with increasing $x$ (Fe,
Cu content). The results are compared with those for corresponding
Zr--base metallic glasses and ion-implanted Hf films.
\end{abstract}


\submitto{\JPCM}

\maketitle

\section{Introduction}
Glassy TE--TL alloys (TE and TL being the early and late
transition metal, respectively) have been extensively studied in
recent decades \cite{R1} and the interest in these alloys further
increased after the discovery of TE--TL base bulk metallic glasses
\cite{R2,R3,R4}. These studies revealed several unusual phenomena
\cite{R5,R6,R7,R8}, which has lead to the development of novel concepts for the
calculation of their properties \cite{R9,R10}. In TE--TL alloys
the composition range for the formation of the amorphus state by rapid quenching from the melt is quite wide and in favourable cases it spans from $20-70$~at.~\% of TL component. Such a broad composition range enables a detailed study of the changes in the electronic band
structure and properties on alloying, through comparison between
the model and experiment \cite{R1,R11,R12}.

In nonmagnetic amorphous TE--TL alloys, several properties which are
related to the electronic density of states (DOS) show simple,
sometimes linear variation with TL content
\cite{R1,R6,R7,R11,R12,R13,R14, R14a,R15}. These simple variations of the properties correlate with
ultraviolet photoemission spectroscopy (UPS) results
for the same alloy systems \cite{R1,R16}, which showed that DOS at the Fermi level ($E_F$), \NE, is
dominated by TE d--states. Accordingly, in amorphous TE--TL alloys, the effect of alloying with TL
can be approximated with the dilution of amorphous TE \cite{R13}.
So far, a majority of results on TE--TL alloys has been obtained
for Zr--TL metallic glasses, rendering a comparison between alloy
systems based on different TE (eg.\ Ti, Zr, Hf) rarely possible
\cite{R11,R12,R17}. This is particularly true for
superconductivity, with only a few results for superconducting
transition temperatures \Tc\ of Ti-- \cite{R18} and Hf--base
\cite{R19,R20} metallic glasses.

Here we report, to our knowledge first, systematic study of
superconductivity in Hf--Fe and Hf--Cu metallic glasses. Our
results show that the variation of \Tc\ with $x$ in \HfFex\ and
\HfCux\ glassy alloys is quite similar to that observed in
corresponding Zr$_{100-x}$Fe$_x$ and Zr$_{100-x}$Cu$_x$ metallic
glasses \cite{R14, R14a,R21}. In particular \Tc\ decreases with
$x$ and the rate of decrease is much faster for $x =$ Fe than for
Cu. The magnitudes of \Tc\ in Hf--base alloys are about two times
lower than those in corresponding Zr--base alloys \cite{R14,
R14a,R21}. The dressed density of states at the Fermi level, \Ng,
also decreases with increasing $x$. Thus, superconductivity in
Hf--base alloys is consistent with ``split--band'' electronic
structure of glassy TE--TL alloys \cite{R1,R16,R19}.

\section{Experimental}
\HfFex\ $(x=20,25)$, \HfCux\ $(x=30,40,50)$, and
Ti$_{65}$Cu$_{35}$ amorphous ribbons were prepared by melt
spinning of master alloys with the predetermined concentration in
either pure Ar \cite{R22} or He \cite{R23} atmosphere. The ribbons
were typically $10\,\mu$m (Hf--Fe) and $20\,\mu$m (all other
alloys) thick and their amorphousness was verified by X--ray
diffraction \cite{R22,R24}. About $9\,$mm long samples for
resistivity measurements were glued by GE--varnish on the sample
holder of a $^3$He cryostat inserted into $16/18\,$T
superconducting magnet. The current and voltage wires were glued
with silver paste onto the samples. The resistivity measurements
were performed by low frequency ($22\,$Hz) ac method with rms
current $I = 0.1\,$mA in the temperature range $0.3\,\mathrm{K} -
290\,$K in magnetic field $B\le 5\,$T, perpendicular to the broad
surface of the ribbon and to the current direction. The
temperature was measured with calibrated Cernox thermometer
situated close to the samples. The resistivity was determined from
the measurements of resistance, length, mass and density of
samples \cite{R13,R14, R14a}. Due to finite width of the silver
paste contacts the uncertainty in the absolute resistivity values
was about 5\%. This uncertainty propagated into the values of the
density of states \Ng. Some data relevant to our samples are given
in table \ref{tbl:TableOne}.
\begin{table}
\caption{\label{tbl:TableOne} Measured and calculated parameters
for Hf--Fe and Hf--Cu metallic glasses. $\rho$ is resistivity at
$2\,$K, $\alpha$ is the temperature coefficient of resistivity,
\Tc\ is the superconducting transition temperature, $\dHdT$ is the
initial slope of the upper critical field, and \Ng\ is the dressed
density of states obtained from \myrefeq{eq:Ngamma}.}
\begin{tabular}{lccccc}
\br Alloy & $\rho$ ($\mu\Omega$cm) & $\alpha$ ($10^{-4}$K$^{-1}$)
& \Tc\ (K) &
            $\dHdT$ & \Ng\ \\

 &  &  &  & (kOe/K) & (states/eV atom) \\

\mr
Hf$_{80}$Fe$_{20}$ & 206 & $-1.2$ & 1.86 & 38 & 2.27 \\
Hf$_{75}$Fe$_{25}$ & 200 & $-2.0$ & 1.12 & 35 & 2.14 \\
Hf$_{70}$Cu$_{30}$ & 206 & $-1.3$ & 1.36 & 28.5 & 1.64 \\
Hf$_{60}$Cu$_{40}$ & 212 & $-1.2$ & 0.82 & 26.1 & 1.40 \\
Hf$_{50}$Cu$_{50}$ & 210 & $-1.1$ & 0.44 & 23.7 & 1.16 \\ \br
\end{tabular}
\end{table}

\section{Results and discussion}
\begin{figure}
\rightline{
\includegraphics*[scale=\myscale]{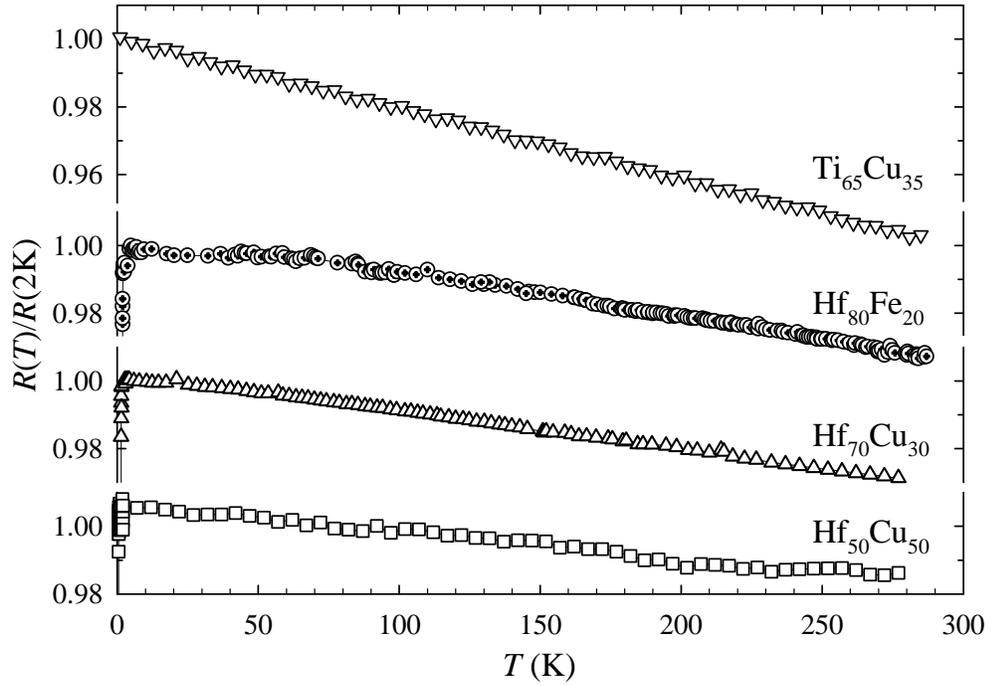}}
\caption{\label{fig:RdT}Temperature dependence of normalized
resistance for representative Hf--based metallic glasses and
\TiCu\ amorphous alloy.}
\end{figure}

Figure \ref{fig:RdT} shows the
variation of resistance with temperature for \TiCu\ and selected
Hf--Cu, Fe glassy alloys. As is usual for glassy
TE$_{100-x}$TL$_x$ alloys with high resistivity ($\rho\ge
140\,\mu\Omega$cm) \cite{R6,R13,R25} all our samples had negative
temperature coefficients of resistivity (TCR). The literature
values for resistivities and TCRs of corresponding Hf--Cu and
\TiCu\ \cite{R6,R25} agree quite well with our results (table
\ref{tbl:TableOne}). In particular, our $T=2\,$K resitivities,
$\rho(2\,\mathrm{K})$, are a few percent higher than the room
temperature resistivites, $\rho(290\,\mathrm{K})$, of other
authors \cite{R6,R25,R26}.

\begin{figure}
\rightline{
\includegraphics*[scale=\myscale]{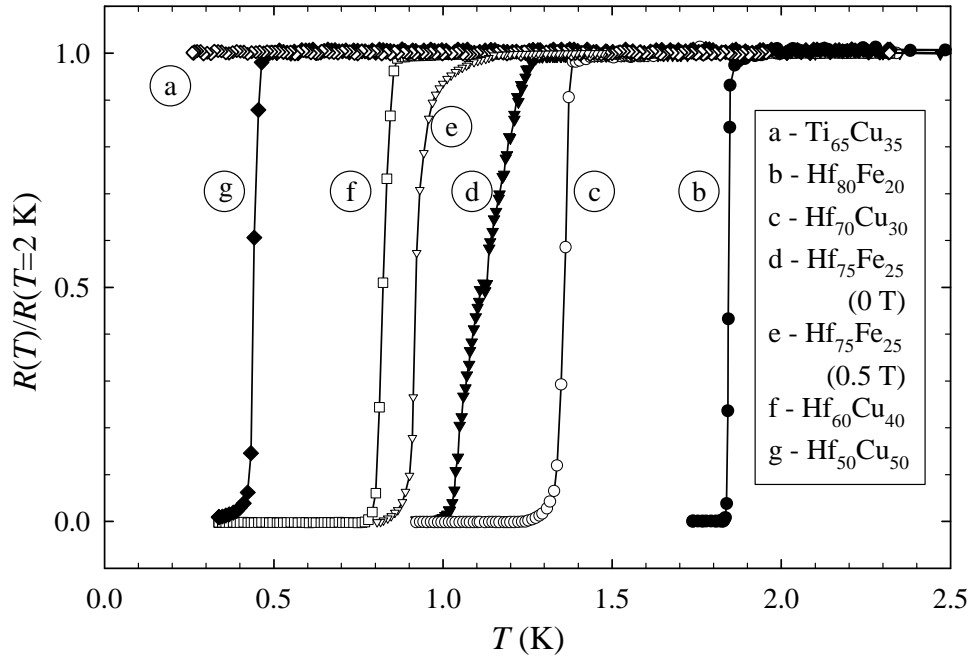}}
\caption{\label{fig:TC}Resistive transitions for Hf--based
metallic glasses. For Hf$_{75}$Fe$_{25}$ alloy transitions curves
in magnetic field 0 and $0.5\,$T are shown.}
\end{figure}

Figure \ref{fig:TC} shows the variations
of normalized resistance $R(T)/R(2\,\mathrm{K})$ with temperature
below $2.5\,$K for all studied alloys. All Hf--Cu,Fe samples
become superconducting within the explored  temperature range
($T\ge 0.3\,$K). Except for Hf$_{75}$Fe$_{25}$ alloy, all other
samples show very narrow superconducting transitions with typical
widths (from 0.1 to $0.9\rho(2\,\mathrm{K})$) $\Delta \mTc \le
0.04\,$K, which can be regarded as an indication of good quality
(homogeneity) of studied samples \cite{R13,R14, R14a,R19,R26}. The
transition width for Hf$_{75}$Fe$_{25}$ alloy, $\Delta \mTc \cong
0.14\,$K, is somewhat larger but not unusual for amorphous alloys.
As illustrated in \myreffig{fig:TC} the resistive transition of
this alloy became narrower in applied field, which allowed
reliable determination of the variation of the upper critical
field with temperature, $\mHcT$, also for this alloy. The values
of superconducting transition temperatures (defined as midpoints
of resistive transitions) are given in table \ref{tbl:TableOne}.
Sample \TiCu\ showed no sign of superconductivity down to
$0.3\,$K, which is consistent with the reported $\mTc \cong
0.06\,$K for this alloy \cite{R18}.

\begin{figure}
\rightline{
\includegraphics*[scale=\myscale]{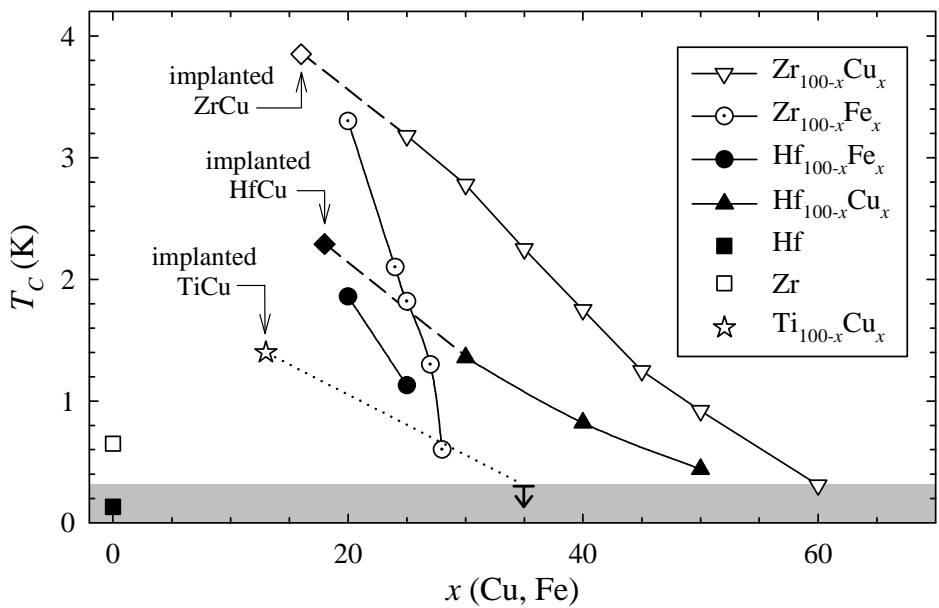}}
\caption{\label{fig:TCx}Superconducting transition temperature
\Tc\ for Hf--base (closed symbols; our work) and Zr--base (open
symbols; \cite{R14}) metallic glasses. Maximum \Tc\ for Cu--ion
implanted Ti (open star), Zr ($\lozenge$), and Hf
($\blacklozenge$) films, \cite{R27} and for crystalline Zr
($\square$) and Hf ($\blacksquare$) \cite{R28} are also shown.
Shaded area denotes temperatures unattainable in our experiment.}
\end{figure}

In \myreffig{fig:TCx} we compare the variations of
zero-field {\Tc}s with concentration $x$ for our \HfFex\ and
\HfCux\ alloys, with the literature results for Zr--Fe and Zr--Cu
alloys \cite{R14, R14a,R21}. Also shown are the maximum {\Tc}s
(i.e.\ the highest attainable \Tc\ in given TE$_{100-x}$Cu$_x$
series) of presumably amorphous Ti--Cu, Hf--Cu and Zr--Cu thin
films, obtained by low temperature ion implantation \cite{R27},
which seem to extrapolate quite well the results for metallic
glasses to lower Cu contents. Since there are no previous results
for \Tc\ of Hf--Fe,Cu glassy alloys, we can compare our results
only with that for splat cooled Hf$_{70}$Ni$_{30}$ foil \cite{R19}
with $\mTc = 1.5\,$K. Judging by the relation between {\Tc}s of
similar Zr--Ni and Zr--Cu alloys \cite{R12,R14, R14a,R21}, \Tc\ of
Hf$_{70}$Ni$_{30}$ alloy \cite{R19} is consistent with $\mTc =
1.36\,$K for our Hf$_{70}$Cu$_{30}$ alloy. The transition
temperatures of pure crystalline (hcp) Zr and Hf
(\myreffig{fig:TCx}) are about an order of magnitude lower than
maximum {\Tc}s of Zr--Cu and Hf--Cu amorphous alloys. This is
qualitatively consistent with the observed \cite{R1,R12,R15,R16}
and calculated \cite{R11,R12} higher \NE\ in dilute amorphous
TE--TL alloys than those of pure crystalline (hcp) TE metals. As
seen from \myreffig{fig:TCx} the variations of \Tc\ with $x$ in
Zr--Fe,Cu and Hf--Fe,Cu amorphous alloys are qualitatively very
similar, the main difference is that {\Tc}s of Hf--Fe,Cu alloys
are about two times lower than those of corresponding Zr--Fe,Cu
alloys. Like in Zr--base alloys the rate of decrease of \Tc\ with
$x$ in Hf--base alloys is much faster for Fe than for Cu alloy.
This is due to onset of magnetic correlations such as the spin
fluctuations and/or formation of magnetic moments/clusters which
cause strong pair-breaking \cite{R14, R14a,R19,R24}.

\begin{figure}
\rightline{
\includegraphics*[scale=\myscale]{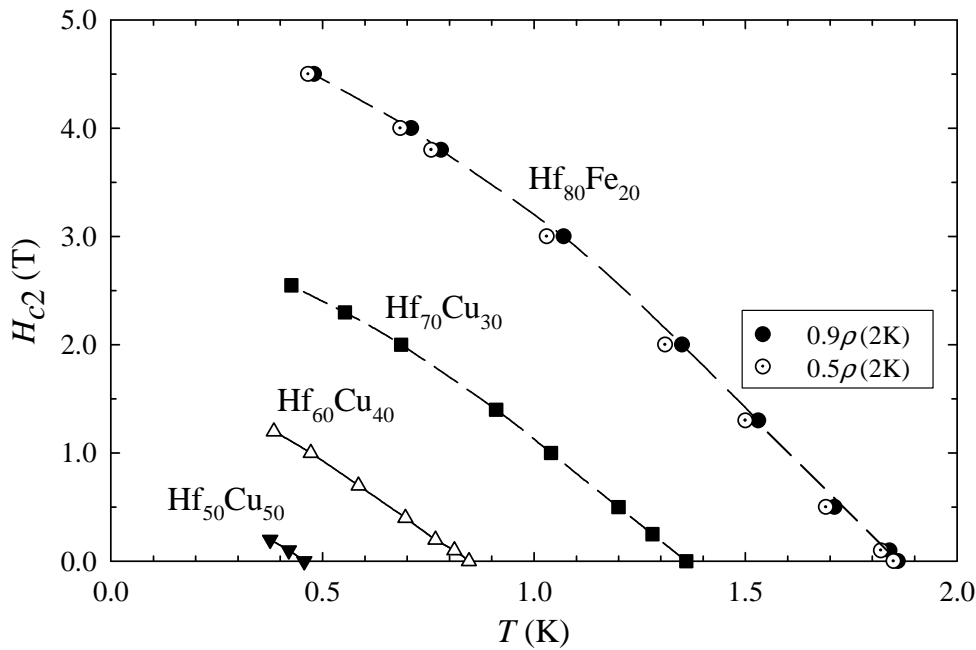}}
\caption{\label{fig:HC2}Upper critical field \Hc\ of
representative Hf--based metallic glasses. See text for definition
of \Hc.}
\end{figure}

Lower {\Tc}s of Hf--Fe,Cu alloys, compared to those of Zr--Fe,Cu,
are consistent with a decrease of \NE\ on going from Zr to Hf (due
to the increase of the bandwidth), but may also be affected
\cite{R28} by the different Debye temperatures of Zr-- and
Hf--base alloys. Unfortunately, there are no measurements of the
low temperature specific heat (LTSH) of Hf--base metallic glasses
\cite{R12} which are necessary in order to explain the difference
between {\Tc}s of Zr--base and Hf--base alloy systems. In the
absence of LTSH, useful information about nature of
superconductivity in metallic glasses can be obtained from the
measurements of upper critical field $\mHcT$ \cite{R14,
R14a,R19,R20,R21,R26,R29}. The variation of \Hc\ with temperature
in TE--TL metallic glasses is usually well described by the Werthamer-Helfand-Hohenberg
theory \cite{R30} and a fit of experimental results to the model
enables one to determine the spin-orbit interaction parameter,
$\lso$, and the Maki paramagnetic limitation parameter $\alpha$
\cite{R31}. However, such fits yield reliable results for the
above parameters (especially $\lso$) only if the measurements
extend to sufficiently low temperature, $T/\mTc\le 0.1$
\cite{R20}. The \HcT\ variations for our Hf--Fe,Cu alloys are
shown in \myreffig{fig:HC2}.
\Hc\ was defined with $0.9\rho(2\,\mathrm{K})$, but -- as
illustrated for Hf$_{80}$Fe$_{20}$ alloy -- the variation of \Hc\
with $T$ for $0.5\rho(2\,\mathrm{K})$ criterion was virtually the
same. Due to low {\Tc}s of alloys we have studied, our
measurements are limited to $T/\mTc\ge0.2$ which is not sufficient
for the accurate estimate of both $\lso$ and $\alpha$. Instead, we
can use rather well defined initial slopes of $\mHc$, $\dHdT$, for
our alloys (table \ref{tbl:TableOne}) in order to estimate their
dressed densities of states, \Ng\ \cite{R14, R14a,R19,R21,R26}.
\Ng\ can be calculated from \cite{R30}:
\begin{equation}\label{eq:Ngamma}
\mNg = -\frac{\pi M}{4k_B N_A \rho d}
        \left( \frac{d\mHc}{dt} \right)_{t=1},
\end{equation}
where $k_B$ is Boltzmann constant, $N_A$ the Avogadro number, $M$
the molecular weight, $d$ the mass density, and $t=T/\mTc$. The
product $\rho d$ can be expressed via resistance $R$, length $l$,
and mass $m$ of the sample, $\rho d=(mR/l^2)$ \cite{R14,
R14a,R26}. The values of \Ng\ for Hf--Cu,Fe alloys calculated by
using \myrefeq{eq:Ngamma} decrease with increasing Fe,Cu content
(table \ref{tbl:TableOne}) in the same fashion as \Ng\ in
corresponding Zr--Fe,Cu alloys \cite{R14, R14a}. However, the
magnitudes of \Ng\ in Hf--base alloys are some 10--12\% lower than
those in corresponding Zr--base alloys \cite{R14, R14a}. As in
Zr--base alloys \cite{R14, R14a,R19} a clear correlation exists
between the values of \Ng\ and \Tc. In metallic glasses the values
of \Ng\ calculated from \myrefeq{eq:Ngamma} usually agree well
with those obtained more directly from the coefficient of a linear
term in LTSH, $\gamma$ \cite{R12,R14, R14a,R19,R20,R21,R26,R29}:
\begin{equation}
\mNg_{\mathrm{LTSH}} = \frac{3\gamma}{\pi^2k_B^2}.
\end{equation}
The dressed density of states is enhanced by many body
interactions in respect to a band (bare) density of states, \NE.
In particular, $\mNg=(1+\lep+\lesf)\mNE$, where $\lep$ and $\lesf$
are the electron-phonon and electron-spin fluctuations interaction
parameters, respectively. Since our Hf--base alloys were
paramagnetic \cite{R24} with magnetic susceptibilities well below
of those for corresponding Zr--base alloys \cite{R14, R14a} we
expect $\lesf$ to be small for Hf--Fe, and negligible for Hf--Cu
alloys. For a reliable estimate of $\lep$ the LTSH measurements
are required \cite{R12}. Since at present no LTSH results for
Hf--base glassy alloys exist \cite{R12} we cannot make accurate
estimates of \NE\ for studied alloys.

In amorphous TE--TL alloys the electron-phonon enhancement factor
can also be estimated from the temperature variation of the
thermopower, $S(T)$ \cite{R6}. Such an estimate for amorphous
Hf$_{50}$Cu$_{50}$ alloy yields $\mNg/\mNE \cong 1.4$, nearly the
same as that obtained for Zr$_{50}$Cu$_{50}$ alloy \cite{R32}.
Another estimate of $\lep$ in Hf--Cu glassy alloys can be obtained
by dividing our results for \Ng\ with the calculated values of
\NE\ for amorphous \HfCux\ alloys \cite{R12,R33}. The ratio
between our \Ng\ and (interpolated) values of \NE\ decreased from
about 1.4 ($x=30$) to $\cong 1.2$ ($x=50$). This calculation gave
\NE\ values for amorphous Hf--Cu alloys about 10\% lower than the
values of \NE\ in corresponding Zr--Cu alloys \cite{R12}.

In principle, $\lep$ can also be obtained from the approximate
proportionality between $\lep$ and \NE\ derived for disordered
transition metal alloys of a given series \cite{R34} which was
found applicable to several 4d (Zr,Mo)-base metallic glasses
\cite{R19}. There, the coefficient of $\lep$ vs.\ \NE\ variation
for 4d and 5d series was found to be quite similar \cite{R34},
whereas that for 3d series was sizeably smaller. This result is in
qualitative agreement with the estimates of $\lep$ from $S(T)$
\cite{R32} for equiatomic Ti--Cu, Zr--Cu and Hf--Cu amorphous
alloys. Thus, the electron-phonon enhancement in Hf--Cu glassy
alloys is probably quite similar to that in Zr--Cu alloys and the
main reason for lower {\Tc}s in the former system may be higher
ionic mass of Hf (lower Debye temperature, $\Theta_D$ \cite{R35})
and lower \NE\ \cite{R12}.

Near absence of superconductivity in Ti--base metallic glasses
\cite{R12,R18}, also confirmed by us (figures \ref{fig:TC} and
\ref{fig:TCx}), is puzzling. Since in these systems both \Ng\ and
$\Theta_D$ are higher than those in corresponding Zr--base and
Hf--base metallic glasses \cite{R12}, an inefficient
electron-phonon coupling is required to explain their low {\Tc}s
\cite{R33}.

\section{Conclusion}
The first systematic study of superconductivity in Hf--based
metallic glasses has been reported. A clear correlation between
the values of \Tc\ and the dressed density of states \Ng\ has been
established. With exception of
Hf$_{75}$Fe$_{25}$ alloy, higher \Ng\ corresponds to higher \Tc.
More rapid suppression of \Tc\ with $x$ in \HfFex\ alloys than in
\HfCux\ is probably caused by magnetic effects. In
general, the variations of \Tc\ and \Ng\ in Hf--Fe,Cu metallic
glasses with Fe,Cu content are qualitatively the same as those in
corresponding Zr--Fe,Cu glassy alloys  which is
consistent with very similar electronic structures of these
alloys. Considerably lower values of \Tc\ in Hf--based metallic
glasses than those in corresponding Zr--based alloys are probably
due to the lower Debye temperatures $\Theta_D$, and electronic
densities of states \NE\ in former system. For a
more detailed insight into the superconductivity of Hf--based
metallic glasses the additional measurements of the low
temperature specific heat (yielding $\Theta_D$) and perhaps
tunnelling experiments (giving more directly electron-phonon
coupling) are required.

\section*{Acknowledgments}

The samples Hf$_{70}$Cu$_{30}$ and Hf$_{60}$Cu$_{40}$ have been
prepared by Dr.\ L.\ Varga and Dr.\ I.\ Bakonyi from Research
Institute for Solid State Physics and Optics, Hungarian Academy of
Sciences. This work was supported by the Croatian Ministry of
Science, Education and Sports projects Nos.\ 119-1191458-1023 and
119-1191458-1019.


\section*{References}

\begin{thebibliography}{10}


\bibitem{R1}
Beck H and G{\"{u}}ntherodt H~J (eds) 1994 {\em Glassy Metals
III\/} ({\em
  Topics in Applied Physics\/} vol~72) (Berlin: Springer)

\bibitem{R2}
Peker A and Johnson W~L 1993 {\em Appl.\ Phys.\ Lett.\/} {\bf 63}
2342

\bibitem{R3}
Das J, Tang M~B, Kim K~B, Theissmann R, Baier F, Wang W~H and
Eckert J 2005
  {\em Phys.\ Rev.\ Lett.\/} {\bf 94} 205501

\bibitem{R4}
Inoue A, Zhang W, Zhang T and Kurosaka K 2001 {\em Acta
Materialia\/} {\bf 49}
  2645

\bibitem{R5}
Gey W, Eschner W and {Yu M Galperin} 1993 {\em Phys.\ Rev.\ B\/}
{\bf 48} 15666

\bibitem{R6}
Howson M~A and Gallagher B~L 1988 {\em Phys.\ Reports\/} {\bf 170}
265

\bibitem{R7}
Marohni{\'{c}} {\v{Z}}, Babi{\'{c}} E, Guberovi{\'{c}} M and
Morgan G 1988 {\em
  J.\ Non--Cryst.\ Solids\/} {\bf 105} 303

\bibitem{R8}
Whang S~H, Polk D~E and Giessen B~C 1982 {\em 4th Int. Conf. On
Rapidly
  Quenched Metals\/} ed Masumoto T and Suzuku K (Sendai) p 1365

\bibitem{R9}
Jank W, {Ch Hausleitner} and Hafner J 1991 {\em Europhys.\
Lett.\/} {\bf 16}
  473

\bibitem{R10}
{Ch Hausleitner} and Hafner J 1990 {\em Phys.\ Rev.\ B\/} {\bf 42}
5863

\bibitem{R11}
Mankovsky S, Bakonyi I and Ebert H 2007 {\em Phys.\ Rev.\ B\/}
{\bf 76} 184405

\bibitem{R12}
Bakonyi I 1995 {\em J.\ Non--Cryst.\ Solids\/} {\bf 180} 131

\bibitem{R13}
Babi{\'{c}} E, Risti{\'{c}} R, Miljak M, Scott M~G and Gregan G
1981 {\em Solid
  State Commun.\/} {\bf 39} 139

\bibitem{R14}
Altounian Z and Strom-Olsen J~O 1983 {\em Phys.\ Rev.\ B\/} {\bf
27} 4149

\bibitem{R14a}
Batalla E, Altounian Z and Strom-Olsen J~O 1985 {\em Phys.\ Rev.\
B\/} {\bf 31}
  577

\bibitem{R15}
Risti{\'{c}} R and Babi{\'{c}} E 2007 {\em Mater.\ Sci.\ Eng.\
A\/} {\bf
  449-451} 569

\bibitem{R16}
Oelhafen P, Hauser E and G{\"{u}}ntherodt H~J 1979 {\em Solid
State Commun.\/}
  {\bf 35} 1017

\bibitem{R17}
Bakonyi I 2005 {\em Acta Materialia\/} {\bf 53} 2509

\bibitem{R18}
Hickey B~J, Greig D and Howson M~A 1986 {\em J.\ Phys.\ F:\ Met.\
Phys.\/} {\bf
  16} L13

\bibitem{R19}
Tenhover M and Johnson W~L 1983 {\em Phys.\ Rev.\ B\/} {\bf 27}
1610

\bibitem{R20}
Nordstr{\"{o}}m A, Dahlborg U and Rapp {\"{O}} 1993 {\em Phys.\
Rev.\ B\/} {\bf
  48} 12866

\bibitem{R21}
Samwer K and v~Lohneysen H 1982 {\em Phys.\ Rev.\ B\/} {\bf 26}
107

\bibitem{R22}
Revesz A, Cziraki A, Lovas A, Padar J, Ledvai J and Bakonyi I 2005
{\em Z.\
  Metallkd\/} {\bf 96} 874

\bibitem{R23}
Babi{\'{c}} E, Butcher S, Day R~K and Dunlop J~B 1985 {\em 5th
Int. Conf. On
  Rapidly Quenched Metals\/} ed Steeb S and Warlimont H (Amsterdam: Elsevier
  Science Publishers B.V.) p 1157

\bibitem{R24}
Paji{\'{c}} D, Zadro K, Risti{\'{c}} R, {\v{Z}}ivkovi{\'{c}} I,
Skoko {\v{Z}}
  and Babi{\'{c}} E 2007 {\em J.\ Phys.:\ Condens.\ Matter\/} {\bf 19} 296207

\bibitem{R25}
Pavuna D 1985 {\em Solid State Commun.\/} {\bf 54} 771

\bibitem{R26}
Karkut M~G and Hake R~R 1983 {\em Phys.\ Rev.\ B\/} {\bf 28} 1396

\bibitem{R27}
Meyer J~D and Stritzker B 1983 {\em Z. Phys. B -- Condensed
Matter\/} {\bf 54}
  25

\bibitem{R28}
Collings E~W and Ho J~C 1971 {\em Phys.\ Rev.\ B\/} {\bf 4} 349

\bibitem{R29}
Poon S~J 1983 {\em Amorphous Metalic Alloys\/} ed Luborsky F
(London:
  Butterworths) p 432

\bibitem{R30}
Werthamer N~R, Helfand E and Hohenberg P~C 1966 {\em Phys.\
Rev.\/} {\bf 147}
  295

\bibitem{R31}
Maki K 1966 {\em Phys. Rev.\/} {\bf 148} 362

\bibitem{R32}
Gallagher B~L and Hickey B~J 1985 {\em J.\ Phys.\ F:\ Met.\
Phys.\/} {\bf 15}
  911

\bibitem{R33}
Cyrot-Lackmann F, Mayou D and {Nguyen Manh} D 1988 {\em Mater.\
Sci.\ Eng.\/}
  {\bf 99} 245

\bibitem{R34}
Dynes R~C and Varma C~M 1976 {\em J.\ Phys.\ F:\ Met.\ Phys.\/}
{\bf 6} L215

\bibitem{R35}
McMillan W~L 1968 {\em Phys.\ Rev.\/} {\bf 157} 331

\end{thebibliography}
\end{document}